\newif\ifAMStwofonts
\AMStwofontstrue

\def\eeth{ E_{_{\rm th}} }

\def\gsim{~\rlap{$>$}{\lower 1.0ex\hbox{$\sim$}}}

\def\ltsim{\lower.5ex\hbox{$\; \buildrel < \over \sim \;$}}
\def\gtsim{\lower.5ex\hbox{$\; \buildrel > \over \sim \;$}}
\def\ltsim{\lower.5ex\hbox{$\; \buildrel < \over \sim \;$}}
\def\gtsim{\lower.5ex\hbox{$\; \buildrel > \over \sim \;$}}

\def\cv{ v}

 \def\ag{A}
  \def\rg{\rho_{_{\rm g}}}

\def\dd{\,{\rm d}}

\def\rs{{r_{_{\rm s}}}}

\newcommand{\etal}{{\it et. al.}\ }
\newcommand{\beq}{\begin{equation}}
\newcommand{\eeq}{\end{equation}}
\def\beqa{\begin{eqnarray}}
\def\eeqa{\end{eqnarray}}
\def\la{\lower.5ex\hbox{$\; \buildrel < \over \sim \;$}}
\def\ga{\lower.5ex\hbox{$\; \buildrel > \over \sim \;$}}
\def\fixit#1{}

\def\dd{{\rm d}}
\def\apj{ApJ}
\def\apjl{ApJL}
\def\mnras{MNRAS}
\def\aap{A\&A}

\documentclass{elsart3p}
\usepackage{ifpdf}
\usepackage{graphicx,natbib,amssymb}

\begin{document}

\begin{frontmatter}
\title{ Gravo-thermodynamics of the Intracluster Medium: negative heat capacity and dilation of cooling time scales }

\author{Adi Nusser}
\address{Physics Department and the Asher Space Science Institute- 
Technion, Haifa 32000, Israel}

\ead{adi@physics.technion.ac.il}

\begin{abstract}
The time scale for cooling of  the gravitationally bound  gaseous  intracluster medium (ICM)  is not determined by radiative processes alone. 
If the  ICM is in quasi-hydrostatic equilibrium in the fixed gravitational field of the dark matter halo then energy losses incurred by the gravitational potential energy of the gas should also be taken into account.
This  ``gravitational heating" has been known for a while using explicit solutions to the equations of motion. 
Here, we re-visit this effect by 
applying the virial theorem to gas in quasi-hydrostatic equilibrium in 
an external gravitational field, neglecting the gravity of the gas.
For a standard NFW form of halo profiles and for a finite gas density, 
the response of the gas temperature to changes in the total energy is 
significantly delayed. 
The effective cooling time could be prolonged by more than an order of magnitude inside the  scale  radius ($\rs$) of the halo. 
Gas lying at a distance twice the scale radius, 
has negative heat capacity so that the temperature
 increases as a result of energy losses. 
Although external heating (e.g. by AGN activity)  is still required  to explain the 
lack of cool ICM near the center,  the analysis here may circumvent
the need for heating in farther out regions where the effective cooling time could be prolonged to become larger than the cluster age and also explains the increase of 
temperature with radius in these regions.  
\end{abstract}

\end{frontmatter}

\maketitle

\section{Introduction}

Clusters of galaxies are the most massive virialized objects 
observed in the Universe. Their potential depths correspond to 
virial temperatures of $1-10\; \rm keV$ $(10^7-10^8 \; \rm K$) and the baryon number density in the inner regions could be as high as $0.1 \rm cm^{-3}$ (e.g. Vikhlinin \etal 2005; Pointecouteau, Arnaud \& Pratt 2005).  For these temperatures and densities, radiative losses are expected to  the  bring the temperature in the central regions down to $\gtsim 10^4 \; \rm K$  
within the available time.  
 Yet in none of the observed clusters does the temperature  drop to the
level dictated by cooling alone.
The absence of significant amounts of cold gas in the cores of massive clusters is 
a major puzzle posed by X-ray observations of massive clusters (e.g. Peterson \etal 2001).
 Hence,   efficient heating mechanisms
must operate at the cores of all cooling clusters.  
The most popular mechanism for suppressing cooling is energy released by 
an  AGN in the central cluster galaxy (cf.  Quilis \etal 2001, 
Babul \etal 2002, Kaiser \& Binney 2003, Dalla Vecchia \etal 2004,
Roychowdhury \etal 2004, Voit \& Donahue 2005, Nipoti \& Binney 2005,
and references therein) or by multiple AGN activity in all galaxies in the cores of clusters (Nusser, Silk \& Babul 2006; Eastman et. al. 2007; Nusser \& Silk 2008).  
Over-pressurized  ejecta from the AGN transform into hot bubbles that 
eventually reach pressure equilibrium with the ICM and proceed to rise buoyantly 
away from the center. 
These bubbles could heat the ICM 
by means of shock waves generated as they expand to reach the ICM pressure (Nusser, Silk \& Babul 2006), and 
by drag forces when they become buoyant (e.g. Churazov et. al. 2001).
  Mechanical activity near the center could also generate  sound
waves which are believed to eventually dissipate their energy in the ICM (Pringle 1989, 
Ruszkowski \etal 2004, Heinz \& Churazov 2005, Fujita \& Suzuki 2005, Sanders \& Fabian 2007).
To balance cooling in a cluster of   X-ray luminosity of $L_{\rm x}\sim 
10^{44}\rm erg \; s^{-1}$, a central AGN must produce $\sim 10^{60} \rm erg$ 
over the entire life-time of the cluster. For the most massive clusters (potential depths
corresponding to velocity dispersions $>500 \; \rm km/s$)
the required heating  could be more than an order of magnitude  larger than 
  the observed range of AGN energy output
in galaxy clusters, based on the $pV$ content of X-ray cavities (e.g. Best \etal 2007). 
This is not too worrying since weak shocks could certainly compensate for the missing energy needed to balance cooling.   
For less massive clusters the $pV $ energy is sufficient to balance cooling (e.g. B\^irzan \etal 2004). 
The challenge, however, is to arrange for  efficient energy transport
from the AGN over the entire cooling core, or  out to distances of up to $\sim 100\;  \rm kpc$.

The temperature in the inner regions increases gradually as 
 we move away from the center. At first, this behavior may seem 
 reasonable since the radiative cooling becomes  more efficient nearer to the  center.  
 But, 
  the cooling time is significantly shorter than the cluster age 
  over a significant part of the inner regions and the ICM had ample opportunity to cool to very low temperatures (e.g. Fig. 12 in Wise, McNamara \& Murray  2004). 
So why is there not a temperature plateau extending over the region where the cooling time is shorter than the cluster age? 
  One explanation might be that,   on account of the lower density, heat conduction is 
more significant as we  move away from the center.  However, 
heat conduction is not universally  important in these regions   
(e.g. Wise, McNamara \& Murray 2004).
Here show that the cooling time could significantly be modified when the potential energy of the ICM in the dark halo is taken into account.  
We will use a version of the virial theorem to show that
the potential energy will absorb some of the  energy loss incurred by the system. In some cases  the potential energy will decrease by an amount larger than the actual loss, forcing the system to compensate the energy difference by increasing  its  thermal energy. This is the case of negative heat capacity.   
This phenomenon is sometimes referred to as gravitational heating 
and has been discussed previously (e.g. Fabian \& Nulsen 1977) and
is evident in numerical simulations of the ICM. 
However, the description in terms of the virial theorem as is done here is new and offers a simple  analysis for assessing the dependence of the effective  cooling time on the assumed halo profile.

\section{The virial theorem}
Hereafter we will assume spherical symmetry
and denote by $r$ the distance from the  center.   
Let $\rg(r)$, $u(r)$,  and $P=(\gamma-1)\rg u$ be, respectively,  the gas density, energy per unit mass,  and pressure, where $\gamma$ is the adiabatic index.
 The   temperature, $T$, is related to $u$ by 
$u=k_{_B} T/(\gamma-1)/m$, where $m$ is the mean particle
mass and $k_{_B}$ is the Boltzman constant.
We assume a  gas obeying the equation
\begin{equation}
\label{eq:he}
\rg g-\frac{\dd P}{\dd r}=0 \; ,
\end{equation}
where  $g$ is the gravitational force field per unit mass. This equation 
is applicable  in quasi-hydrostatic equilibrium so that 
the acceleration of the gas is negligible.  
Multiplying (\ref{eq:he}) by $r$ and integrating over the volume from $r=0$ to $R_0$ gives
the virial theorem, 
\begin{equation}
\label{eq:vir}
3 (\bar P -P_0)V+W=0
\end{equation}
where $ V=(4\pi/3) R_0^3$, $P_0=P(R_0)$ is the external pressure, $ \bar P=  4\pi \int_0^{R_0}
\dd r r^2 P(r)/V$  is the average pressure inside $R_0$,
and 
the gravitational term, $W$,  is 
\begin{equation}
W=4\pi \int_0^{R_0} r^3 \rg g(r)  \dd r 
\end{equation}
 A more general derivation which includes gas motions could be found in  Ostriker \& McKee (1988). 
The energy of the system in the volume $V$ is written as the sum of the thermal energy $\bar P V/(\gamma-1)\propto N k_{_{\rm B}} T  $ ($N$ is the total number of particles) and the gravitational potential energy, $U$, 
\begin{equation}
\label{eq:ene}
E=U+\frac{\bar P V}{\gamma -1}\; 
\end{equation}
where 
\begin{equation}
U=4\pi \int_0^{R_0} \rg \Phi r^2 \dd r\; ,
\end{equation}
and the system is assumed to  reside in a static external gravitational
potential $\Phi$ and neglected gravity of the gas.

From the virial theorem (\ref{eq:vir}) and the energy equation (\ref{eq:ene}) we obtain global relations between infinitesimal variations (denoted by the prefix $\delta$ ) in the total energy, $E$, the thermal energy $\eeth$, $V$, $W$ and $U$. 
Keeping a constant external pressure $P_0$ these 
relations are 
\begin{equation}
\label{eq:dvir}
\delta W=3P_0 \delta V -3(\gamma-1)\delta \eeth \; , 
\end{equation}
and
\begin{equation}
\label{eq:dene}
\delta E=\delta U +\delta \eeth\; , 
\end{equation}
where we have  used the expression $\eeth=\bar P V/(\gamma-1)$ for the thermal energy. These relations must hold for any change in the state of the system.
For radiative losses, the energy loss in time 
$\delta t$ is  $\delta E=n_{_{\rm e}}\Lambda(T)\delta t $
where $ n_{_{\rm e}}$ is the electron number density and $\Lambda$ is the cooling rate. 
Even if this energy is extracted initially from 
the thermal part, $\eeth$, subsequent evolution of the system will establish the relations  (\ref{eq:dvir}) and (\ref{eq:dene}). We are working under the assumption of quasi-hydrostatic equilibrium so that 
any bulk motions generated during this process are
neglected. In any case,  if dissipation is important 
then  significant gas motions will be converted  into heat, restoring the above relations. 
 
We are set now to derive a relation between $\delta E$ and $\delta \eeth$.
We write  $\delta V=(\delta V/\delta W)\delta W $ in the virial relation (\ref{eq:dvir}) 
to obtain,
\begin{equation}
\label{eq:dw}
\delta W=\frac{3 (\gamma-1)}{3P_0\frac{\delta V }{\delta W}-1} \delta \eeth \; .
\end{equation}
Writing $\delta U=(\delta U/\delta W)\delta W$ and $\delta V=(\delta V/\delta W) \delta W$ in the relation (\ref{eq:dene})  while taking   $\delta W$ from 
 (\ref{eq:dw}) we get
\begin{equation}
\label{eq:cth}
\delta E={\cal C}\delta \eeth\; ,
\end{equation}
where 
\begin{equation}
\label{eq:C}
{\cal C}=1+\frac{\delta U}{\delta W}\frac{3(\gamma-1)}{ 3P_0\frac{\delta V}{\delta W}-1}\; .
\end{equation}
The quantity ${\cal C}$ gives the ratio of the heat capacity to the  standard thermodynamic heat capacity computed without gravity. 
Hence we call 
${\cal C}$ the relative heat capacity (RHC).
The sign  of ${\cal C}$ determines whether the thermal energy, $\eeth$,  and hence the temperature, $T\propto \eeth/ N$,  will increase or decrease as a result of  changing  the total energy, $E$.  If ${\cal C}<0$ holds, then the heat capacity is negative, i.e. the temperature  increases when we extract energy from the system. 
For $P_0=0$, the condition  ${\cal C}<0$
implies 
\begin{equation}
\label{eq:Ccond}
\frac{\delta U}{\delta W}>\frac{1}{3(\gamma-1)} \; .
\end{equation}
For positive  RHC, ${\cal C} >0$, the 
response time of the gas temperature to variations in its energy   is prolonged 
by a factor $\cal C$.  
For example,  the effective  cooling time is $ {\cal C} t_{_{\rm cool}} $ where $t_{_{\rm cool}} 
\sim k_{_{\rm B}} T/(n_{_{\rm e}}\Lambda)$
is the usual radiative cooling time. 

\section{applications to various forms of halo gravitational potentials}

We begin with 
the calculation of the RHC, $\cal C$, for
 power-law potentials of the form, $\Phi=\ag/r^n$ so that $g=\ag n/r^{n+1}$, where $n\ne 0$ and $\ag$ are constants. The constant $\ag$ is negative for $n>0$ and positive otherwise.
In this case $W=n U$ and $\delta U/\delta W=1/n$ and for $P_0=0$ 
we have
\begin{equation}
\label{eq:Cn}
{\cal C}=1-\frac{3}{n}(\gamma-1)\; .
\end{equation}
Thus $\cal C$ is negative for
\begin{equation}
0<n<3(\gamma-1)\; ,
\end{equation}
which gives $0<n<2$ for  $\gamma=5/3$.
To estimate the RHS, $\cal C$, for a non-vanishing external pressure, $P_0$, 
we need the quantity $ \delta W/\delta V$ which  depends on gas density profile, $\rg$, in the system. 
We work here with a power-law density profile of the form,
$\rg=B/r^\alpha$ and we compute  $\delta W/\delta V$ 
under variations of  the external radius $R_0$ assuming that the index $\alpha$ and the total mass, $M_0$,  inside $R_0$ remain constant.  
Since $M=4\pi\int r^2 B/r^\alpha \dd r $ we get 
\begin{equation}
B =\frac{3-\alpha}{4\pi}M R_0^{\alpha-3}\; . 
\end{equation}
Evaluating $W $ we get
\begin{equation}
W=  \frac{n(3-\alpha)}{3-\alpha-n} \ag M  {R_0^{-n}}\; .
\end{equation}
Therefore,
\begin{equation}
\frac{\delta W} {\delta V}=- 
\frac{n^2(3-\alpha)}{3-\alpha-n} \frac{\ag M }{4\pi R_0^{3+n}}
\end{equation}
and
\begin{equation}
\label{eq:dvdw}
\frac{\delta V} {\delta W}=-\frac{3 V}{n W}\; .
\end{equation}
Since $W<0$,  this quantity is negative for $n<0$ and,  as seen in (\ref{eq:C}), the heat capacity is positive for any $P_0$. 
For $n>0$, ${\delta V}/ {\delta W}$ is  positive. Thus,  an inspection of 
(\ref{eq:C}) reveals, for $\delta U/\delta W>0$ the existence of external pressure  could  result in  ${\cal C }<0$ even if ${\cal C} >0$ for $P_0=0$. For this to happen, the value of $P_0$ has to be adjusted such that $3P_0 \delta V/\delta W-1 $ is small and
 negative. 
Substituting $W$ from  the virial theorem (\ref{eq:vir})
into (\ref{eq:dvdw}) and using the later into (\ref{eq:C}) we get the RHC in terms of the 
pressure ratio, $\bar P/P_0$, as follows
 \begin{equation}
 {\cal C}=1-\frac{3}{n}\frac{\gamma-1}{1-\frac{1}{n(\bar P/P_0-1)}}\; .
\end{equation}
This expression reduces to (\ref{eq:Cn}) for $\bar P/P_0\gg 1$.

  An intriguing  case is  $\Phi=A\ln r$ 
 and $g=-A/r$ ($A>0$). By the requirement of constant mass inside the varying radius $R_0$ we find that $W$ is constant.
  Therefore, $\delta U/\delta W$ is either $-\infty$ or $+\infty$ depending on the sign of $\delta U$. The sign of $\delta U$ is sensitive to the assumed density form of the gas. For example, for  $\rho=B/r^\alpha$ as above we get 
  $ U=const$ so that $\delta U=0$. For a mass distribution 
  confined to a shell of negligible thickness, $\delta U $ is
  positive when the shell is brought closer to the center and negative otherwise.

\begin{figure*} 
\centering
\includegraphics[height=0.5\textheight,angle=90]{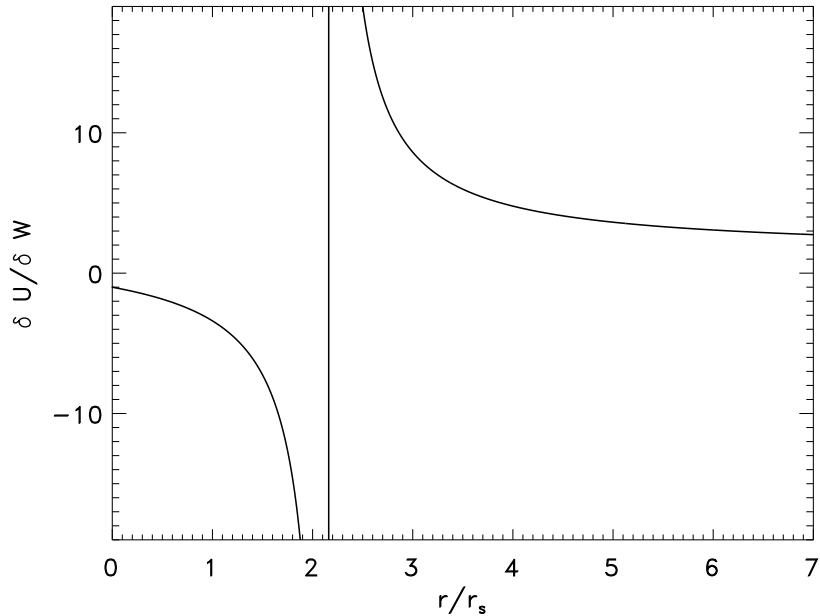}
\vspace{10pt}
\caption{The quantity $\delta u/\delta w
=[\dd \Phi/\dd r]/[\dd (r g)/\dd r]  $ corresponding to an infinitesimal displacement, $\delta r$, of  a fluid element  as a function of  position.}
\label{fig:duw1}
\end{figure*}

We now consider the implications of the relation (\ref{eq:cth}) 
for realistic distributions of dark matter in halos. 
Therefore, we adopt  the parametric  form proposed by  Navarro, Frenk \& White (1996) (hereafter NFW) for the density profile, which is motivated by N-body simulations and is consistent with the distribution of dark matter in observed clusters (Pointecouteau,  Arnaud \&  Pratt 2005). 
The  gravitational potential and force field
for a halo following the NFW profile are
\begin{equation}
\Phi(r)=-\frac{G M_{\rm v}}{r} \frac{f}{c^2} \ln (1+ c s)
\end{equation}
and
\begin{equation}
g(r)=-\frac{\dd \Phi}{\dd r}=-\frac{G M_{\rm v}}{r^2} \frac{f}{c^2} \left[\ln (1+ c s)-\frac{c s}{c s
+1}\right] \; .
\end{equation}
where $c$ is the concentration paramater, $s=r/R_{\rm v}$ is the distance from
the halo center in units of the virial radius $R_{\rm v}$, $M_{\rm v}$ is the virial mass of the halo, and $f= c^2/ [\ln(1+c)-c/(1+c)]$. The virial mass is related to the virial radius by 
$M_{\rm v}=(4\pi/3) 200 \rho_{_{\rm crit}}R_{\rm v}$ where $\rho_{_{\rm crit}}$ is the critical 
cosmic density. 
 The structure of a halo is, therefore,  determined uniquely by the $c$ and $R_{\rm v}$.
 The scale radius $r_{_{\rm s}}\equiv  c  R_{\rm v}$  marks the transition 
 from $g=const $ near $r=0$ to $g\propto 1/r$ as we move further out. 
For this (non power-law) form of the gravitational field, the quantity $\delta U/\delta W$ depends on the assumed form of the variation in the 
gas distribution which is determined by 
energy gains and losses.
We find it most instructive to focus on  effects of local density variations on $\cal C$.
Therefore, we present here $\delta u/\delta w$ resulting from infinitesimal displacements, $\delta r$, of  a fluid element of a given mass as a function of its position, i.e., 
$\delta u/\delta w=\delta \Phi/\delta (r g)$. 
This type of variation  is relevant for cooling/heating processes in a 
shell of  finite thickness lying at a distance $\sim r$ from the center. 
We  show, 
in figure \ref{fig:duw1}, $\delta u/\delta w$ as a function of the radius in units of 
the scale radius, $r_{_{\rm s}}$. There a singularity at   $ r=r_{_{\rm th}}\sim 2.15 r_{_{\rm s}}$ which corresponds to a vanishing $ \delta W$. For $r>r_{_{\rm th}} $,
the quantity $\delta U/\delta W$ is positive so that according to (\ref{eq:C}) the RHC, $\cal C$, is negative 
(for $P_0=0$). As we move to the inner regions at  $r<r_{_{\rm th}}$ the quantity 
$\delta U/\delta W$ switches signs and so the RHC becomes positive. However, 
the RHC, $\cal C$,  is  significantly larger than unity. At $r=0$, ${\cal C}=3 $ for $P_0=0$ and $\gamma=5/3$ so that the cooling time is prolonged by a factor of 3.
This is a modest boosting in the cooling time since the standard radiative cooling time scale could as short as 0.01 of the life-time of clusters. 
However, the dilation of cooling time scales  in farther out regions  
   could be large enough so as to exceed the cluster age. 
The RHC, $\cal C$, is also affected by   $P_0 \delta V/\delta w$ evaluated at the 
outer boundary of the shell.
If the average pressure $\bar P$ inside the shell is large compared to $P_0$
then this term could be neglected. For realistic clusters the pressure is a steep function of radius for a significant  part of the inner cluster cores.
Of course the external pressure at the inner
boundary of the shell should also be considered.  However, we assume that the 
inner shell radius is small compared to the 
outer radius so that the work done by the the pressure on the inner shell is small.

\section{Concluding Remarks}
\label{sec:Conclude}

We used a simplified model of the ICM to study the 
its gravo-thermodynamical properties. 
In quasi-hydrostatic equilibrium, the inclusion of the change in the 
potential energy prolongs the response of the   gas temperature  in the inner regions lying 
within $r_{_{\rm th}}\approx  2 r_{_{\rm s}}=2R_{_{\rm v}}/c$. 
Outside this radius, the form of dark halo gravitational potential is such that  the temperature is increased as a result of energy loss, i.e. the gas heat capacity is negative. 
The boosting of the cooling time in the inner region formally diverges   
   at $r_{_{\rm th}}$ reaching  a  factor of $3$ as
at  $r=0$.

Our results may circumvent the need for heating the ICM in regions where the 
standard radiative cooling time is an order of magnitude shorter than the 
life-time of the cluster.  Those regions lie at a significant fraction of $r_{_{\rm s}}$
where  the prolonged effective cooling time could be larger than  the cluster age.
as is the case for example for the cluster A1068. For this cluster, $r_{_{\rm s}}\sim 400 \; \rm kpc $ (Pointecouteau, Arnaud \& Pratt 2005) and  the ratio of  standard radiative cooling time to the cluster age  is  $\sim 0.1- 1$ over the region 
between $\sim 70 \; \rm kpc$ to $\sim 300\;  \rm kpc$. 
Our results do not eliminate the need for heating of the ICM 
in central regions ($r<< r_{_{\rm th}} $) since the effective cooling time there is still 
shorter than the cluster age.

 As mentioned before, the approach taken here aims at addressing a specific point related to the heat capacity of the ICM. The analytic methods used  here could be followed only by invoking a simplified (perhaps oversimplified) of the ICM. 
The effect of gravitational heating has appeared in a variety of 
forms in the literature. For example, it has been discussed by  Fabian \& Nulsen (1977) using specific solutions to subsonic collapse of gas in a
$1/r$ gravitational potential. It is also seen in the one dimensional
simulations by Omukai \& Nishi (1998) of self-gravitating gas for the formation of primordial protostellar clouds. 
It is  also evident in the solutions for self-similar cooling flows by Bertschinger (1989). 
However, as is shown here, analysis of these problems using virial theorem offers a simple way to understand the physical effect
and its relation to the halo profile and external pressure.

\section{Acknowledgments}
The author  acknowledges useful discussions
with Noam Soker and Zac Myers. 
This work is supported by the German-Israeli Foundation for 
Research and Development and by the Asher Space Research
Institute. 


\begin{thebibliography}{}

\bibitem[Allen et al.(2008)]{2008MNRAS.383..879A} Allen, S.~W., Rapetti, 
D.~A., Schmidt, R.~W., Ebeling, H., Morris, R.~G., 
\& Fabian, A.~C.\ 2008, \mnras, 383, 879 

\bibitem[B{\^i}rzan et al.(2004)]{2004ApJ...607..800B} B{\^i}rzan, L., 
Rafferty, D.~A., McNamara, B.~R., Wise, M.~W., 
\& Nulsen, P.~E.~J.\ 2004, \apj, 607, 800 


\bibitem[Babul et al.(2002)]{2002MNRAS.330..329B} Babul, A., Balogh, M.~L., 
Lewis, G.~F., \& Poole, G.~B.\ 2002, \mnras, 330, 329 


\bibitem[Best et al.(2007)]{2007MNRAS.379..894B} Best, P.~N., von der 
Linden, A., Kauffmann, G., Heckman, T.~M., 
\& Kaiser, C.~R.\ 2007, \mnras, 379, 894 

\bibitem[Bertschinger(1989)]{1989ApJ...340..666B} Bertschinger, E.\ 1989, 
\apj, 340, 666 

\bibitem[Churazov et al.(2001)]{2001ApJ...554..261C} Churazov, E., 
Br{\"u}ggen, M., Kaiser, C.~R., B{\"o}hringer, H., 
\& Forman, W.\ 2001, \apj, 554, 261 


\bibitem[Dalla Vecchia et al.(2004)]{2004MNRAS.355..995D} Dalla Vecchia, 
C., Bower, R.~G., Theuns, T., Balogh, M.~L., Mazzotta, P., 
\& Frenk, C.~S.\ 2004, \mnras, 355, 995 


\bibitem[Eastman et al.(2007)]{2007ApJ...664L...9E} Eastman, J., Martini, 
P., Sivakoff, G., Kelson, D.~D., Mulchaey, J.~S., 
\& Tran, K.-V.\ 2007, \apjl, 664, L9 

\bibitem[Fabian 
\& Nulsen(1977)]{1977MNRAS.180..479F} Fabian, A.~C., \& Nulsen, P.~E.~J.\ 1977, \mnras, 180, 479 




\bibitem[Fujita 
\& Suzuki(2005)]{2005ApJ...630L...1F} Fujita, Y., \& Suzuki, T.~K.\ 2005, \apjl, 630, L1 


\bibitem[Heinz 
\& Churazov(2005)]{2005ApJ...634L.141H} Heinz, S., \& Churazov, E.\ 2005, \apjl, 634, L141 


\bibitem[Navarro et al.(1997)]{1997ApJ...490..493N} Navarro, J.~F., Frenk, 
C.~S., \& White, S.~D.~M.\ 1997, \apj, 490, 493 


\bibitem[Nipoti 
\& Binney(2005)]{2005MNRAS.361..428N} Nipoti, C., \& Binney, J.\ 2005, \mnras, 361, 428 


\bibitem[Nusser 
\& Silk(2008)]{2008MNRAS.386.1013N} Nusser, A., \& Silk, J.\ 2008, \mnras, 386, 1013 


\bibitem[Nusser et al.(2006)]{2006MNRAS.373..739N} Nusser, A., Silk, J., 
\& Babul, A.\ 2006, \mnras, 373, 739 

\bibitem[Omukai 
\& Nishi(1998)]{1998ApJ...508..141O} Omukai, K., \& Nishi, R.\ 1998, \apj, 508, 141 



\bibitem[Ostriker 
\& McKee(1988)]{1988RvMP...60....1O} Ostriker, J.~P., \& McKee, C.~F.\ 1988, Reviews of Modern Physics, 60, 1 


\bibitem[Peterson et 
al.(2001)]{2001A&A...365L.104P} Peterson, J.~R., et al.\ 2001, \aap, 365, L104 


\bibitem[Pointecouteau et 
al.(2005)]{2005A&A...435....1P} Pointecouteau, E., Arnaud, M., \& Pratt, G.~W.\ 2005, \aap, 435, 1 


\bibitem[Pringle(1989)]{1989MNRAS.239..479P} Pringle, J.~E.\ 1989, \mnras, 
239, 479 


\bibitem[Quilis et al.(2001)]{2001MNRAS.328.1091Q} Quilis, V., Bower, 
R.~G., \& Balogh, M.~L.\ 2001, \mnras, 328, 1091 


\bibitem[Reynolds et al.(2005)]{2005MNRAS.357..242R} Reynolds, C.~S., 
McKernan, B., Fabian, A.~C., Stone, J.~M., 
\& Vernaleo, J.~C.\ 2005, \mnras, 357, 242 


\bibitem[Roychowdhury et al.(2004)]{2004ApJ...615..681R} Roychowdhury, S., 
Ruszkowski, M., Nath, B.~B., \& Begelman, M.~C.\ 2004, \apj, 615, 681 


\bibitem[Ruszkowski et al.(2004)]{2004ApJ...611..158R} Ruszkowski, M., 
Br{\"u}ggen, M., \& Begelman, M.~C.\ 2004, \apj, 611, 158 


\bibitem[Sanders 
\& Fabian(2007)]{2007MNRAS.381.1381S} Sanders, J.~S., \& Fabian, A.~C.\ 2007, \mnras, 381, 1381 


\bibitem[Vikhlinin et al.(2005)]{2005ApJ...628..655V} Vikhlinin, A., 
Markevitch, M., Murray, S.~S., Jones, C., Forman, W., 
\& Van Speybroeck, L.\ 2005, \apj, 628, 655 


\bibitem[Voit 
\& Donahue(2005)]{2005ApJ...634..955V} Voit, G.~M., \& Donahue, M.\ 2005, \apj, 634, 955 




\end{thebibliography}
\end{document}